# Short-term effect of hyperbaric exposure on Ventilation: A Control Study of 12m-depth Single No-decompression Dive Experiment

Hua CHENG[1]

Lingnan normal university

Author Contribution:

(1) The proposal and design of the research propositions
(2) Conducting tests or surveys;
(3) Obtaining, providing and analyzing data;
(4) Revision of the drafts or final revision of the thesis.

# Abstract

**Objective:** To study to what extent or durations of ventilation effect in a single no-decompression dive of 12 meters to a diver.

**Methods:** There are 29 healthy volunteers divers assigned into SCUBA diving of 12m-depth underwater (the Experimental Group, EG) and chamber dive under 2.2 ATA for 20min (the Control Group, CG) matched with the factors of the *age*, *gender*, *BMI* and *Forced Vital Capacity* (*FVC*). Ventilation function*s* were measured by spirometer before diving and in 1h and 24h of post-hyperbaric exposure. Used independent samples *T* tests to compare the differences between the EG and CG. Analyzed of variance through repeated measurement data of different time point before or after high pressure exposure by *SPSS* 20.0.

**Results:** The *Inspiratory Reserve Volume(IRV)* rises while the *Expiratory Reserve Volume(ERV)* falls significantly in 1h after high pressure release($p<0.05$). So as with the *Inspiratory Capacity (IC)* and the *Vital Capacity (VC)* increased accordingly. The *Ratio of $FEV_{1.0}$ to VC* ($FEV_{1.0}\%t$) is higher in CG than EG ($t$=-2.189, $p$=0.033) due to the change of *VC*. But the effects did not last for 24 h after high pressure relief.

**Conclusions:** Ventilation is restricted during the 20min of hyperbaric exposure whether under 12m-depth water or in a 2.2ATA hyperbaric chamber. But the effect recovered close to normal within 24 h. But the effect recovered close to normal within 24 h. The extent of restriction of underwater diving is larger than the dry air hyperbaric chamber dive. Higher water medium density, submerged compressing blood volume of lower limbs and raising inertia added by portable underwater breathing apparatus all might be attributable to the ventilation effects.

# 1. Introduction

Both immersion and the increasing of pressure/depth in diving environment have likely increase the demand for the respiratory system [1-3]. Because static lung loading grows as immersion and breathing gas density increases with depth, the ventilation system is more important in the environment of SCUBA diving than the terrestrial environment [4]. As is well known for all, blood flow to the contracting skeletal muscles during exercise provided oxygen consumption came from the air through respiratory movement [5]. Hyperbaric circumstances are underwater, with intrathoracic pressures fluctuate drastically and exercise-induced fatigue of respiratory muscle can limit cardiac output and therewith the leg blood flow [6]. Meanwhile, reduction in ventilation volumes further brings down the arterial oxygen content and greatly increased the work of the respiratory muscular[7-9]. Weiterhin, constrictive effect exert on the heart since expansion of lung volume and chest squeeze can quickly change in systolic and diastolic of cardiac function and affect the heart rate and stroke volume as well[10]. This is a complex result formed by numerous molecular interaction effects of stacking based on the adjustability of respiratory reflex and ultimately caused tissue hypoxia [11]. The above mentioned all together will further intensify the occurrence of exercise-induced fatigue and endurance performance obstacles.

Ventilation limitation induced exertion dyspnea and exercise intolerance is an important clinical problem unsolved in exercise as far as today is concerned [12]. Even though the diving technology had been improved and the apparatus had been upgraded all along, SCUBA diving still couldn't escape from the challenge of ventilation limitation. High pressure effects impose on thoracic, pleura or respiratory tract tissue leads to limitation of pulmonary alveoli expansion underwater. And the lung capacity reduction increased the external respiratory work and insufficient in ventilation and/or hypoxemia. Ventilation restriction varies greatly resting with the pattern and time-histories of hyperbaric exposure. But the identifiable parameters for reference that indicated relationships between level of ventilation restriction and depth of the underwater are still rare.

Taking healthy diving trainees as subjects for observation, we carried out a series of experiments grouping on the basis of controlling of depth and speed of decompression in water and a hyperbaric chamber. Relevant parameters of pulmonary ventilation were observed before and after hyperbaric exposure, to aim at the respiratory physiology for diving pressure to provide experimental data for further research.

### Abbreviation

| Terminology | Abbreviation |
|---|---|
| Tidal Volume | $TV$ |
| Inspiratory Reserve Volume | $IRV$ |
| Expiratory Reserve Volume | $ERV$ |
| Residual Volume | $RV$ |
| Inspiratory Capacity | $IC$ |
| Vital Capacity | $VC$ |
| Function Residual Capacity | $FRC$ |
| Total Lung Capacity | $TLC$ |
| Minute Ventilation | $MV$ |
| Maximal Voluntary Ventilation | $MVV$ |
| Forced Vital Capacity | $FVC$ |
| Forced Expiratory Volume in one second | $FEV_{1.0}$ |
| Ratio of FEV1 to FVC | $FEV_{1.0}/FVC\ (FEV_{1.0}\%)$ |
| Ratio of FEV1 to VC | $FEV_{1.0}/VC\ (FEV_{1.0}\%t)$ |
| Forced Expiratory Flow | $FEF_{25\sim75}\%$ |
| Peak Expiratory Flow | *PEF* |
| Forced Expiratory Flow after 25％ of the FVC has been exhaled | *FEF$_{25}$％* |
| Forced Expiratory Flow after 50％ of the FVC has been exhaled | *FEF$_{50}$％* |
| Forced Expiratory Flow after 75％ of the FVC has been exhaled | *FEF$_{75}$％* |

## 2. Methodology

### 2.1 Volunteer

Enrolled volunteers were divided into SCUBA diving group(the Experimental Group, EG) and a chamber diving (the Control Group, CG) matched with the *Age*, *Gender*, *Body Mass Index* (*BMI*) and *Forced Vital Capacity* (*FVC*) respectively. Participated volunteers should be physically healthy and normal in ECG testing, without suffering from an acute respiratory infection in a recent week. They also should be proficiency in SCUBA diving techniques with more than 1 year in diving experience whose diving depth's at least 20m below surface for at least 5 min. Although the experimental settings are safe for every participant, but still all volunteers were informed by informed consents the possible risks and avoiding accidents during the experiment before started. Those cannot manage to complete the whole process will be ruled out by the experiment. Basic physical information parameters (*Date of Birth, Gender, Height, Weight, BMI*) and respiratory function index (*TV, VC, MV, MVV*) detected by the experimenter.

Determination of *Age* and *BMI* is like the following formula (1) -(2).

$$Age(y) = [testing\ Date\ (yy/mm/dd) - birth\ Date(yy/mm/dd)]/365 \quad (1)$$

$$BMI = \frac{Weith(Kg)}{Height(cm)^2} \quad (2)$$

### 2.2 Experimental protocols

Experiment is phased in three temporal durations, namely pre-hyperbaric exposure, 1 h and 24h post-hyperbaric exposure. Subjects' baseline of pulmonary ventilation detection completed before high pressure exposed. Parameters of pulmonary ventilation at 1 h and 24h post-hyperbaric exposure detection should proceed with group by group.

Volunteers were required low-fat, high-protein diet a week before experiment and pulmonary ventilation detection finished 2 days before experiment and pulmonary ventilation detection finished 2 days before exposure. For more accurate and stable data of pulmonary ventilation, one subject measuring need to repeat two more times at one time on the same one experimental condition. Take any necessary average as the experimental data. Similarly here & after.

Acting in accordance with *the Age, Gender, BMI, FVC* of team members, subjects were divided into diving and chamber diving group. The pulmonary function of ventilation testing accomplished in 1 h and 24h after hyperbaric exposure.

Field experimentation for diving group was in a 12-meter deep diving tower where the bottom temperature is 17 degrees centigrade and surface temperature is 23 degrees centigrade. Put on diving suits，with flippers and self-contained breathing apparatus, divers to submergence to the bottom within 2 minutes and ascend to surface in 2 minutes of every single dive bottom time is 20 minutes without doing strenuous exercises. Acquired data of parameters of pulmonary ventilation function by Spirometer at 1 h and 24 h after surfacing.

Through pressurizing atmospheric pressure (100 kPa) to 220kPa within 2 min in a hyperbaric oxygen chamber, chamber diving group stay under pressure of 220kPa for 20 min. Then reduced pressure to atmospheric pressure at full speed, imitates the no-decompression diving process. Compression chamber's temperature controlled in the range of 24 to 28 ±2 degrees centigrade. Data acquisition of pulmonary ventilation functions in 1 h and 24 h after descending.

### 2.3 Measuring method

Pulmonary function of ventilation *VC，FVC，MV& MVV*）was measured sequentially by the Spirometer（MINATO,AS-505）. The accuracy of capacity for the instrument is ±3% or ±50 *mL*, which its flow range is 0~14*L/S* and the precision of flow is ±3% of quantitative value or ±0.01 *L/S*.

It can analyze and diagnosis normal, restrictive, obstructive, combined disturbance of ventilation and make the diagnosis of small airway function according to the velocity of flow capacity of the loop curve.

Participants take standing positions as instructed. With filter and mouthpiece plugged in before the test and make sure not leaking when exhale, the participants learned to hold their coughing while testing and breathe only through the mouth to get more accurate measurements. Their basic information about the *Gender*, *Age*, *Height* & *Weight* of the volunteers should be imputed to differentiate individual participants.

Instructor asked the volunteer subjects to inhale or exhale in sequence in strict accordance with operating instructions procedures of *VC, FVC, MV* and *MVV*. There are parameters such as *TV, IRV, ERV, IC, FEV$_{1.0}$, FEV$_{1.0}$%, PEF, FEF$_{25-75}$, MEF$_{75}$, MEF$_{50}$, MEF$_{25}$* are calculated. *TV, IRV* and *ERV* are 3 basic important components of static lung capacity. Their relationships with *IC* and *VC* are shown in *equation* (3) ~ (4). According to the predictive logistic equation of **Baldwin**, the *VC* and *MVV* prediction values are determined by their *Age*, *height* and *body surface area* (*BSA*), as *equation*(5) ~(8). *MV* is not only influenced by *TV*, but also by breathing rate (*RR*), as *equation* (9). Forced expiratory volume (*FEV, FEV$_{1.0}$, FEV$_{1.0}$%* and *FEV$_{1.0}$%t*) and the expiratory flow (*PEF, FEF$_{25-75}$, MEF$_{75}$, MEF$_{50}$* and *MEF$_{25}$*) constitute important factors in the detection of dynamic lung capacity. Expiratory volume changes over time in exhalation. Their relationships are listed by *equation* (10) ~ (11). Test results data collected and accept the average of the results are the outcome except the values of *MVV*, which only if the variation is lower than 8% in continuous testing should be accepted and take the maximum value recognized as the test results.

$$VC(L) = TV(L) + IRV(L) + ERV(L) \tag{3}$$

$$IC(L) = TV(L) + IRV(L) \tag{4}$$

$$VC_{male}(mL) = \{27.63 - [0.112 \times Age(y)]\} \times Height(cm) \tag{5}$$

$$VC_{female}(mL) = \{21.78 - [0.101 \times Age(y)]\} \times Height(cm) \tag{6}$$

$$MVV_{male}(L) = \{86.4 - [0.522 \times Age(y)]\} \times BSA(m^2) \tag{7}$$

$$MVV_{female}(L) = \{71.3 - [0.474 \times Age(y)]\} \times BSA(m^2) \tag{8}$$

$$MV(L) = TV(L) \times RR(bpm) \tag{9}$$

$$FEV_{1.0}\% = \frac{FEV_{1.0}(L)}{FVC(L)} \times 100\% \tag{10}$$

$$FEV_{1.0}\%t = \frac{FEV_{1.0}(L)}{VC(L)} \times 100\% \tag{11}$$

**2.4 Data statistics**

Differences comparing between the two groups of normal distribution measurement index applied as independent samples *t* test. Pulmonary ventilation indicators of pre- & post- hyperbaric exposure at different time were analysis of variance for repeated data. Run normality test and homogeneity test of variance *t* test apply for the normal distribution data, and nonparametric test is for non-normal distribution, small sample or unequal variances. Apply repeated measurement data for *Mauchly's* Test of *Sphericity* if the results' concomitant probability $p > 0.05$, then the assumption is satisfied and no need to correction. But if *p* acuities were 0.05, the degree of freedom has to be corrected by $\varepsilon$ correction coefficient and accept the *Greenhouse - Geisser* correction results in this experiment. A *p* value of< 0.05 was regarded as statistically significant. Data was analysis by *IBM SPSS* 20.0 for statistics.

## 3. Results

### 3.1 Overall situations

Overall subjects' average *Age* was 22.07 ±1.13 years, average *Weight* 63.5±7.78Kg, an average of *Height* 172±6 cm, average *BMI* at21.4±2.00. There were 12 divers in diving group (Experiment Group, EG, the same below) and 17 divers in chamber diving group (Control Group, CG, the same below) completed the respiratory function test for baseline reference (supporting *Tab*.1). Two-independent sample *K-S* test 2 groups of *Weight*, *Height and BMI* and *FVC* variables for the normality. Results demonstrate that *Z* value is 0.701 ~ 0.937and *p* is higher than 0.05, significant level. Therefore believe that the variables in the two groups are approximately normal distribution (supporting *Tab*.2).

Equal variances assumed of *p* higher than the significance level of 0.05 in a test of homogeneity of variances of *Age*, *Weight*, *Height*, *BMI* and *FVC* in two groups. And *t* test results suggest the variables above are no significant differences because of *p* higher than the significance level of 0.05(supporting *Tab*.3).

The mean values of Static lung volume (namely *TV*, *IRV*, *ERV*, *IC* and *VC*) and time-related lung volume (*FVC*, $FEV_{1.0}$, $FEV_{1.0}/FVE\%$, $FEV_{1.0}/VC\%$, *PEF*, $FEF_{25-75}$, $MEF_{75}$, $MEF_{50}$, $MEF_{25}$, and *MV*) analyzed as shown in *Tab.1*.

### 3.2 Grouping situation
### 3.2.1 Static lung volume

Although the variation range of *TV* before and after hyperbaric exposure is not significant ($F=0.258$, $p=0.0773$). Variables of *TV* raised in 1 h ($t=-0.831$, $p=0.413$) and 24 h ($t=-0.040$, $p=0.969$) post-hyperbaric exposure. Yet there was no significant difference between two groups ($t=0.390$, $p=0.698$), such as *Tab.2 & Fig.1-A*.

Significant differences were found in *IRV* different time pre- & post- hyperbaric exposure ($F=3.787$, $p=0.029$). *IRV* is depending on the integrative action of thoracic elastic resistance and inspiratory muscle strength. It suggested antagonism of high pressure exposure to thoracic elastic resistance and inspiratory muscle strength is still strong in 1 h ($t=-3.356$, $p=0.002$), much less in 24 h after exposure ($t=0.773$, $p=0.446$). There was no significant difference between two groups ($t=0.318$, $p=0.751$), such as *Tab.3 & Fig.1-B*.

*ERV* declines in 1 h ($t=2.298$, $p=0.029$) and 24 h ($t=0.829$, $p=0.414$) after hyperbaric exposure. *ERV* reflects gas reserve capacity of the lung. It is decided by the rises of diaphragm, thoracic elastic resistance and bronchioles obstruction as exhale forcefully. And the founding indicates gas reserve capacity of the lung is shrinking after hyperbaric exposure ($F=4.910$, $p=0.011$). There was no significant difference between two groups ($F=0.825$, $p=0.444$), such as *Tab.4 & Fig.1-C*.

Influenced by the *IRV* and *TV*, a significant difference present in *IC* in different times being after hyperbaric exposure ($F=8.085$, $p=0.000$). Similar to *IRV*, *IC* significantly increased because of high-pressure effect last for a short term of 1h ($t=-3.589$, $p=0.001$) and decline in 24 h ($t=0.821$, $p=0.419$) after exposure. There was no significant difference between two groups ($F=1.979$, $p=0.148$), such as *Tab.5 & Fig.1-D*.

*VC* is the sum of the *TV*, *IRV* and *ERV*. Under the comprehensive function of many factors, *VC* change significantly ($F=4.078$, $p=0.022$) because of high-pressure effect, while increased ($t=-2.638$, $p=0.013$) in 1 h and declined in 24 h ($t=2.759$, $p=0.010$) significantly post-hyperbaric exposure. There was no significant difference between two groups ($F=1.003$, $p=0.373$), such as *Tab.6 & Fig.1-E*.

### 3.2.2 Time-related lung volume

Made comparison between EG & CG and found no significant difference in *FVC* before or after high pressure exposure ($t=0.000$, $p=0.998$) (supporting *Tab.4*). $FEV_{1.0}$ ($t=-1.579$, $p=0.118$) and $FEV_{1.0}\%$ ($t=-1.771$, $p=0.080$) in EG is higher than in the CG but there was no statistically significant difference between them(supporting Tab.5-6), as in *Fig.2-A*、*Fig.2-B*、*Fig.2C*. But we found out that $FEV_{1.0}/VC$ in CG is significantly higher than EG ($t=-2.189$, $p=0.033$), as in *Tab.7 & Fig 2-D*.

Fluctuating value of *PEF* in two groups before and after hyperbaric exposure is not significant ($F=0.069$, $p=0.933$), which indicated that large bronchus obstruct by high pressure effect was not significant(supporting *Tab.2*), as in *Fig.2-F*. Similarly in the same way with $FEF_{25-75}$ ($F=0.185$, $p=0.832$), $MEF_{75}$ ($F=0.012$, $p=0.988$), $MEF_{50}$ ($F=0.240$, $p=0.787$)、$MEF_{25}$ ($F=0.502$, $p=0.608$), suggested that bronchioles obstruct were not significant as well(supporting *Tab.*8-11), as in *Fig.2-E*、*2-F*.

The experimental results suggest that similar changes happened in $FVC$, $FEV_{1.0}$ and $FEV_{1.0}/FVC\%$ between 2 groups, which proved the blockage effect of high-pressure exposed in two groups were not significant. And $FEV_{1.0}/VC$ is considerably greater in CG than in EG, because of $VC$ is far lower in CG than the EG. Under pressure of 2.2 ATA, all values of $PEF$, $FEF_{25-75}$, $MEF_{75}$, $MEF_{50}$, $MEF_{25}$ are higher in CG, which indicated the degree of restricting effect for the airway in EG is higher.

$MVV$ declined in EG members while a rise in the CG after exposure ($t = 0.327$, $p = 0.746$). But the change is not statistically significant. It can be thought of no statistical significance of $MVV$ before and after hyperbaric exposure (supporting *Tab*.12), such as in *Fig.2-H*.

The value of $MV$ is influenced by $TV$ and $RR$. It is normally less when resting, but increases while exercise. It rises to 40 ~ 60$L$ during moderate-intensity exercise. Both $MV$ and $RR$ decline after exposure, but the differences are not statistical significance because of concomitant probability less than 0.05. The results indicated that $MV$ and $RR$ do not change significantly in 12 m/2.2 ATA high pressure exposures(supporting *Tab*.13-14), such as in *Fig.2-G*.

## 4. Discussion

When submerging down underwater, the ambient pressure of surroundings raised, the breathing gas density and partial pressure, external respiratory work and pulmonary physiological dead space are likewise increased[3, 13]. These factors might likely have an effect on the respiratory function. Such as gas density and partial pressure caused respiratory resistance increased, extra respiratory work motivated respiratory muscle fatigue and low ventilation perfusion ($\dot{V}_A/\dot{Q}$) Ratio, all leading to hypoventilation and $CO_2$ retention [14, 15]. There is supposed to be very large amounts of gas bubbles generated after no-decompression air dives even if divers obey standardized diving protocols. High bubble loads have closely relationship with the VGE travelling to the circulation. Even so, there is still no acute decompression-related pathology was observed [16]. Because it is believed that no considerable impacts on $\dot{V}_A/\dot{Q}$ Ratio after dives within the no-decompression-stop limits [17]. So our study only concentrates on ventilation of external respiration and irrespective of gas exchange or internal respiration regulation in this section.

Lung structure and pulmonary function change significantly along with the age increase. With the degenerate of lung elastic elements, loss of the parenchymal tissue, dilation of alveolar ducts and bronchioles, decreases of chest wall compliance, reduction of the intercostal muscle mass and force, ventilation volume expands and gas exchange surface lessens [18]. In addition to age, ventilation is correlated with the type of exercise, *Gender*, *Height*, *Weight*, and *BMI*, as well [19]. In the study, we couldn't observe too much fluctuation on $TV$ before or after hyperbaric exposure, because the study design settings are multi-index matching was concerned, such as the *Age*, *Sex*, *Weight*, *Height* and *FVC*. Increasing inspiratory muscles' strength overcomes the resistance from the thoracic with the pressure from the hydrostatic pressure/hyperbaric air environment and to complete respiratory movement. The reason why $IRV$ increased in 1h after exposure because the greater inertia produced by the negative pressure increasing inside the lung through the thoracic expansion and the elastic load of chest wall augment which caused by forced air inhalation. As higher gas density and pressure increased airways resistance, exhalation gas flow also reduced. Meantime tissue elastic load of lung increase, so as negative pressure within lungs, which contribute to the decline of involuntary movements of air-breathing. The increase of pulmonary elastic load is responsible for the reduction in transmural pressure, and result in the occurrence of decreasing of $ERV$. This result is consistent with previous cognition on $EVR$ during immersion, which is the subject breathing harder trying to increase the diameter of the airway and reducing the airways resistance [20-22].

In addition, vascular contraction due to the drops in ambient temperature results in reduction in pulmonary perfusion. While diving underwater, higher pulmonary artery pressure and bigger vascular volume during exercise make residual volume ($RV$) increase and $VC$ decreased [23]. The stress of increased static lung load affects the lung volume at the end of the expiratory [24]. In that case, the length of respiratory muscle couldn't reach the optimum length and couldn't maintain enough power to increase the ability of breathing [25, 26][3].

When airflow obstruct, forced expiratory prolong, the $FEV_{1.0}$ and $FEV_{1.0}/FVC\%$ reduced. When ventilation is restricted, the compliance of the lung and thorax reduced. Therefore $VC$ decreases. The vast majority of *VC* exhales in a very short time lead to the increase of $FEV_{1.0}/VC$ [27-30]. $FEV_{1.0}/VC$ increased in EG after hyperbaric exposure suggested there is more restrictive effect of thoracic activity in EG is higher individual than CG. Because even though the equivalence of pressure underwater or in the hyperbaric chamber , the higher medium density of water produce higher hydrostatic pressure than the air in the chamber, plus the close-fitting diving suit, and the weight of breathing apparatus itself has strain on the respiratory system.

Peak expiratory flow（*PEF*）is resting with the respiratory muscle strength of individuals and the presence of airway obstruction[31]. Reduced in $FEF_{25-75}$, $MEF_{50}$ or $MEF_{25}$ has been noted as a symbol of obstruction in small airways. Studies had proved that respiratory system inertia is usually proportional to the gas density increases under standard atmospheric pressure (1ATA). Respiratory resistance originates in inside is mainly the increasing gas density and quality, which add up systematic inertia of respiratory.

Minute ventilation volume (*MV*) didn't change much pre- or post- hyperbaric exposure between 2 groups. But we discovered the value of *MVV* is far lower than the predicted value. That is due to the fact that not only the gas density makes the airway resistance rise underwater, but also the lung elastic load increase due to the pulmonary blood volume and oxygen partial pressure rose triggered by immersion. That submerged generate additional mechanical load to the chest wall, making static pressure load across the chest wall as well as lung compliance lower. Therefore the airway resistance and lung elastic load can lead to the reduction of ventilation underwater. Moreover, more physiological dead space and die cavity/tidal volume ratio（$V_d/V_t$）affected the gas diffusion and ventilation distribution in lungs[20].

## 5. Conclusion

Ventilation is restricted during the 20min of hyperbaric exposure whether in 12m-depth underwater or 2.2ATA hyperbaric chamber. Volume of forced inspiratory increase while volume of expiratory decrease in physiologically rectifies ventilation in 1h after high pressure effect removed. But the ventilation malfunctioning removed and recovered close to normal within 24 h. Results suggested that restrictions of high pressure mainly retard exhalation, which enlarging residual volume and then lessen the pulmonary elasticity. Ultimately result in increased of small airway's resistance. Extent of restriction of underwater is larger than dry air hyperbaric chamber. It may be attributable to the higher water medium density, submerged compressing blood volume of lower limbs and raising inertia added by portable underwater breathing apparatus.

The results show that the fluctuations of pressure in the diving exercise might result in the change of the pulmonary ventilation capacity, and the result is universality. The particularity and uniqueness of this job are mainly concerned with the short-term effects of high pressure exposure rather than long-term effects. The effects of gas exchanging on pulmonary ventilation at 12m/2.2ATA for 20min have not yet been considered in this study. There are not enough valid data for comparative between experimental group and control group during hyperbaric exposure for technical reason, which is the issues to be further studied.

# ACKNOWLEDGEMENTS

Special thanks are for Zhanjiang Diving School that provides experimental area and the main experimental equipment. And also like to thank the diving instructors from Zhanjiang Diving School who had offered me a lot of help and in the experiment.

This study is funded by①Science Research Project of Lingnan Normal University 《Study on the Relationship between the Asymptomatic VGE and VA/Q in single no-decompression dive》（ZL1508）;②Research Program of Science and Technology of Zhanjiang《Pretreatment of HBO vs. NBO for Intervention of VGE in Air Diving: a Randomized Double-blind Controlled Study》（2015B01115）.

# SUPPORTING INFORMASTION

Supporting Tab. 1  General information of study participants

| Group | Gender （Male=1，Female=2） | Age （y） | Weight （Kg） | Height （cm） | BMI | FVC （L） |
|---|---|---|---|---|---|---|
| Experimental Group =1 | 1 | 22 | 67 | 170 | 23.18 | 3.75 |
| | 1 | 22 | 78 | 172 | 26.37 | 3.30 |
| | 1 | 23 | 68 | 175 | 22.2 | 3.43 |
| | 1 | 20 | 65 | 173 | 21.72 | 4.59 |
| | 1 | 21 | 69 | 179 | 21.53 | 4.09 |
| | 1 | 22 | 58 | 170 | 20.07 | 3.22 |
| | 2 | 23 | 53 | 166 | 19.23 | 3.18 |
| | 1 | 21 | 67 | 179 | 20.91 | 4.22 |
| | 1 | 21 | 67.2 | 176 | 21.69 | 3.73 |
| | 1 | 23 | 70 | 179 | 21.85 | 4.11 |
| | 1 | 23 | 60 | 165 | 22.04 | 3.72 |
| | 2 | 22 | 53.7 | 165 | 19.72 | 2.69 |
| | 1 | 23 | 63 | 172 | 21.3 | 3.68 |
| | 1 | 24 | 74 | 186 | 21.51 | 4.55 |
| | 1 | 22 | 74 | 188 | 20.94 | 3.88 |
| | 1 | 23 | 65 | 172 | 21.97 | 3.73 |
| Control Group =2 | 1 | 21 | 57 | 169 | 19.96 | 3.17 |
| | 1 | 23 | 55.5 | 166 | 20.14 | 3.43 |
| | 1 | 20 | 63 | 170 | 21.8 | 3.50 |
| | 1 | 22 | 64 | 176 | 20.66 | 4.43 |
| | 1 | 22 | 74.6 | 171 | 25.51 | 3.50 |
| | 1 | 24 | 63 | 164 | 23.42 | 3.40 |
| | 1 | 21 | 68 | 173 | 22.72 | 3.30 |
| | 1 | 24 | 47 | 163 | 17.69 | 3.47 |
| | 1 | 21 | 58 | 173 | 19.38 | 3.04 |
| | 1 | 23 | 63 | 182 | 19.02 | 4.55 |
| | 2 | 22 | 46 | 163 | 17.31 | 2.90 |
| | 1 | 21 | 68 | 170 | 23.53 | 3.76 |
| | 1 | 21 | 62.2 | 168 | 22.17 | 3.63 |

Supporting Tab. 2  2 groups K-S test for the normality of variables

|  |  | Age (y) | Weight(Kg) | Height(cm) | BMI | FVC (L) |
|---|---|---|---|---|---|---|
| Most extreme difference | absolute value | 0.176 | 0.314 | 0.181 | 0.255 | 0.230 |
|  | negative | 0.020 | 0.314 | 0.181 | 0.255 | 0.230 |
|  | positive | -0.176 | -0.093 | -0.176 | -0.093 | -0.098 |
| K-S  Z |  | 0.468 | 0.832 | 0.481 | 0.676 | 0.611 |
| *p* |  | 0.981 | 0.493 | 0.975 | 0.751 | 0.849 |

Supporting Tab. 3  Test of homogeneity of variances

| | | Age (y) | | Weight(Kg) | | Height(cm) | | BMI | | FVC (L) | |
|---|---|---|---|---|---|---|---|---|---|---|---|
| Group | | 1 | 2 | 1 | 2 | 1 | 2 | 1 | 2 | 1 | 2 |
| $n$ | | 12 | 17 | 12 | 17 | 12 | 17 | 12 | 17 | 12 | 17 |
| $\bar{\chi}$ | | 21.92 | 22.18 | 64.66 | 62.67 | 1.72 | 1.72 | 21.71 | 21.12 | 3.67 | 3.64 |
| $s$ | | 1.00 | 1.24 | 7.24 | 8.26 | 0.05 | 0.07 | 1.85 | 2.12 | 0.53 | 0.49 |
| *Levene* test of variance equations | $F$ | 1.21 | | 0.02 | | 0.33 | | 0.80 | | 0.25 | |
| | $p$ | 0.28 | | 0.89 | | 0.57 | | 0.38 | | 0.62 | |
| $T$ test of mean equation | $t$ | -0.63 | | 0.67 | | 0.14 | | 0.78 | | 0.14 | |
| | $p$ | 0.54 | | 0.51 | | 0.89 | | 0.44 | | 0.89 | |

Supporting Tab. 4  Changes in *FVC* pre- and post- hyperbaric exposure between 2 groups

| Group | FVC ($\bar{X} \pm SD$) | | | sum | F | p |
|---|---|---|---|---|---|---|
| | Pre- | 1h post- | 24h post- | | | |
| EG | 3.67±0.53 | 3.58±0.51 | 3.60±0.46 | 3.62±0.49 | 0.103 | 0.903 |
| CG | 3.64±0.49 | 3.65±0.58 | 3.55±0.50 | 3.62±0.52 | 0.194 | 0.825 |
| sum | 3.65±0.50 | 3.62±0.55 | 3.57±0.47 | 3.62±0.50 | 1.881@ | 0.162@ |
| t | 0.141 | -0.336 | 0.238 | 0.000 | 1.049# | 0.357# |
| p | 0.889 | 0.739 | 0.814 | 0.998 | | |

@：The *F* statistic and *P* values of main effect ；  #：The *F* statistic and *P* values of interaction effect.
\*：*P*<0.05，the difference was statistically significant；  \*\*：*P*<0.01，the difference was significant statistical significance.

Supporting Tab. 5 Changes in $FEV_{1.0}$ pre- and post- hyperbaric exposure between 2 groups

| Group | $FEV_{1.0}$ ($\bar{X} \pm SD$) | | | sum | F | p |
| --- | --- | --- | --- | --- | --- | --- |
| | Pre- | 1h post- | 24h post- | | | |
| EG | 2.79±0.95 | 2.85±0.59 | 2.80±0.73 | 2.81±0.75 | 0.023 | 0.977 |
| CG | 3.07±0.56 | 3.08±0.52 | 2.94±0.59 | 3.03±0.55 | 0.316 | 0.731 |
| sum | 2.95±0.74 | 2.99±0.55 | 2.88±0.65 | 2.94±0.65 | 0.393@ | 0.593@ |
| t | -0.930 | -1.109 | -0.589 | -1.579 | 0.220# | 0.711# |
| p | 0.366 | 0.277 | 0.561 | 0.118 | | |

@: The F statistic and P values of main effect ; #: The F statistic and P values of interaction effect.
*: $P<0.05$, the difference was statistically significant; **: $P<0.01$, the difference was significant statistical significance.

Supporting Tab. 6 Changes in $FEV_{1.0}\%$ pre- and post- hyperbaric exposure between 2 groups

| Group | $FEV_{1.0}\%$ ($\bar{X} \pm SD$) | | | sum | F | p |
| --- | --- | --- | --- | --- | --- | --- |
| | Pre- | 1h post- | 24h post- | | | |
| EG | 75.55%±21.92% | 80.65%±16.96% | 78.43%±19.48% | 78.21%±19.11% | 0.205 | 0.816 |
| CG | 84.55%±12.93% | 84.92%±11.16% | 82.95%±12.75% | 84.14%±12.09% | 0.124 | 0.884 |
| *sum* | 80.82%±17.46% | 83.16%±13.74% | 81.08%±15.72% | 81.69%±15.56% | 0.631@ | 0.474 |
| *t* | -1.390 | -0.820 | -0.756 | -1.771 | 0.544# | 0.513# |
| *p* | 0.176 | 0.419 | 0.456 | 0.080 | | |

@：The *F* statistic and *P* values of main effect ； #：The *F* statistic and *P* values of interaction effect.
\*：*P*<0.05，the difference was statistically significant；\*\*：*P*<0.01，the difference was significant statistical significance.

Supporting Tab. 7  Changes in *PEF* pre- and post- hyperbaric exposure between 2 groups

| Group | PEF ($\bar{X} \pm SD$) | | | sum | F | p |
|---|---|---|---|---|---|---|
| | Pre- | 1h post- | 24h post- | | | |
| EG | 4.17±2.10 | 4.41±1.97 | 4.56±2.02 | 4.38±1.98 | 0.113 | 0.893 |
| CG | 4.76±1.47 | 4.36±1.19 | 4.36±1.57 | 4.49±1.41 | 0.462 | 0.633 |
| sum | 4.52±1.75 | 4.38±1.53 | 4.44±1.74 | 4.45±1.66 | 0.069@ | 0.933@ |
| t | -0.899 | 0.083 | 0.305 | -0.297 | 1.446# | 0.245# |
| p | 0.377 | 0.935 | 0.762 | 0.767 | | |

@: The *F* statistic and *P* values of main effect ; #: The *F* statistic and *P* values of interaction effect.
*: $P<0.05$, the difference was statistically significant; **: $P<0.01$, the difference was significant statistical significance.

Supporting Tab. 8  Changes in $FEF_{25-75}$ pre- and post- hyperbaric exposure between 2 groups

| Group | $FEF_{25-75}$ ($\bar{X} \pm SD$) | | | sum | F | p |
| --- | --- | --- | --- | --- | --- | --- |
| | Pre- | 1h post- | 24h post- | | | |
| EG | 3.13±1.54 | 3.22±1.38 | 3.18±1.46 | 3.18±1.42 | 0.010 | 0.990 |
| CG | 3.48±0.97 | 3.36±0.96 | 3.23±1.14 | 3.36±1.01 | 0.255 | 0.776 |
| sum | 3.34±1.23 | 3.30±1.13 | 3.21±1.26 | 3.28±1.19 | 0.185@ | 0.832@ |
| t | -0.754 | -0.325 | -0.117 | -0.701 | 0.352# | 0.705# |
| p | 0.457 | 0.748 | 0.908 | 0.485 | | |

@：The F statistic and P values of main effect ； #：The F statistic and P values of interaction effect.
*：$P<0.05$，the difference was statistically significant；**：$P<0.01$，the difference was significant statistical significance.

Supporting Tab. 9 Changes in $MEF_{75}$ pre- and post- hyperbaric exposure between 2 groups

| Group | $MEF_{75}$ ($\bar{X} \pm SD$) | | | sum | F | p |
| --- | --- | --- | --- | --- | --- | --- |
| | Pre- | 1h post- | 24h post- | | | |
| EG | 3.88±2.06 | 4.17±1.97 | 4.28±2.04 | 4.11±1.97 | 0.124 | 0.884 |
| CG | 4.54±1.45 | 4.20±1.17 | 4.17±1.56 | 4.30±1.38 | 0.364 | 0.697 |
| sum | 4.27±1.72 | 4.19±1.52 | 4.21±1.74 | 4.22±1.64 | 0.012@ | 0.988@ |
| t | -1.017 | -0.050 | 0.157 | -0.511 | 1.352# | 0.267# |
| p | 0.318 | 0.960 | 0.877 | 0.611 | | |

@: The F statistic and P values of main effect ; #: The F statistic and P values of interaction effect.
*: $P<0.05$, the difference was statistically significant; **: $P<0.01$, the difference was significant statistical significance.

Supporting Tab. 10 Changes in $MEF_{50}$ pre- and post- hyperbaric exposure between 2 groups

| Group | $MEF_{50}$ ($\bar{X} \pm SD$) | | | sum | F | p |
|---|---|---|---|---|---|---|
| | Pre- | 1h post- | 24h post- | | | |
| EG | 3.37±1.71 | 3.32±1.28 | 3.43±1.62 | 3.37±1.50 | 0.015 | 0.985 |
| CG | 3.73±1.04 | 3.57±0.97 | 3.43±1.14 | 3.58±1.03 | 0.364 | 0.697 |
| sum | 3.58±1.34 | 3.47±1.14 | 3.43±1.33 | 3.49±1.25 | 0.240@ | 0.787@ |
| t | -0.725 | -0.601 | 0.003 | -0.711 | 0.501# | 0.608# |
| p | 0.475 | 0.553 | 0.998 | 0.480 | | |

@：The *F* statistic and *P* values of main effect ；#：The *F* statistic and *P* values of interaction effect.
\*：*P*<0.05，the difference was statistically significant；\*\*：*P*<0.01，the difference was significant statistical significance.

Supporting Tab. 11  Changes in $MEF_{25}$ pre- and post- hyperbaric exposure between 2 groups

| Group | $MEF_{25}$ ($\bar{X} \pm SD$) | | | sum | F | p |
| --- | --- | --- | --- | --- | --- | --- |
| | Pre- | 1h post- | 24h post- | | | |
| EG | 2.14±0.98 | 2.16±0.95 | 2.09±0.77 | 2.13±0.88 | 0.017 | 0.983 |
| CG | 2.45±0.64 | 2.44±0.84 | 2.30±0.87 | 2.40±0.78 | 0.204 | 0.816 |
| sum | 2.32±0.80 | 2.33±0.88 | 2.21±0.82 | 2.29±0.83 | 0.502@ | 0.608@ |
| t | -1.055 | -0.844 | -0.650 | -1.498 | 0.121# | 0.886# |
| p | 0.301 | 0.406 | 0.521 | 0.138 | | |

@：The F statistic and P values of main effect ；#：The F statistic and P values of interaction effect.
*：$P<0.05$，the difference was statistically significant；**：$P<0.01$，the difference was significant statistical significance.

Supporting Tab. 12  Changes in *MVV* pre- and post- hyperbaric exposure between 2 groups

| Group | MVV ($\bar{X} \pm SD$) | | | sum | F | p |
|---|---|---|---|---|---|---|
| | Pre- | 1h post- | 24h post- | | | |
| EG | 60.83±24.79 | 59.17±15.56 | 54.07±13.14 | 58.02±18.22 | 0.434 | 0.651 |
| CG | 56.55±17.58 | 55.12±16.96 | 57.94±20.59 | 56.54±18.11 | 0.100 | 0.905 |
| *sum* | 58.32±20.56 | 56.79±16.24 | 56.34±17.72 | 57.15±18.07 | 0.571@ | 0.568@ |
| *t* | 0.545 | 0.655 | -0.573 | 0.375 | 1.691# | 0.194# |
| *p* | 0.590 | 0.518 | 0.571 | 0.708 | | |

@：The *F* statistic and *P* values of main effect ；  #：The *F* statistic and *P* values of interaction effect.
*：$P<0.05$，the difference was statistically significant；**：$P<0.01$，the difference was significant statistical significance

Supporting Tab. 13  Changes in *MV* pre- and post- hyperbaric exposure between 2 groups

| Group | *MV* ($\bar{X} \pm SD$) | | | sum | F | p |
| --- | --- | --- | --- | --- | --- | --- |
| | Pre- | 1h post- | 24h post- | | | |
| EG | 30.50±15.08 | 28.32±13.57 | 28.41±14.87 | 29.08±14.14 | 0.087 | 0.917 |
| CG | 26.95±10.18 | 26.15±10.10 | 23.92±8.34 | 25.67±9.47 | 0.456 | 0.636 |
| *sum* | 28.42±12.32 | 27.05±11.48 | 25.78±11.47 | 27.08±11.68 | 2.985@ | 0.059@ |
| *t* | 0.760 | 0.493 | 1.041 | 1.259 | 0.622# | 0.541# |
| *p* | 0.454 | 0.626 | 0.307 | 0.213 | | |

@: The *F* statistic and *P* values of main effect ; #: The *F* statistic and *P* values of interaction effect.
*: $P<0.05$, the difference was statistically significant; **: $P<0.01$, the difference was significant statistical significance.

Supporting Tab. 14  Changes in *RR* pre- and post- hyperbaric exposure between 2 groups

| Group | RR ($\bar{X} \pm SD$) | | | sum | F | p |
| --- | --- | --- | --- | --- | --- | --- |
| | Pre- | 1h post- | 24h post- | | | |
| EG | 21.51±7.92 | 20.00±7.92 | 20.28±8.35 | 20.60±7.86 | 0.120 | 0.887 |
| CG | 20.92±7.79 | 20.72±10.76 | 16.97±6.18 | 19.54±8.49 | 1.177 | 0.317 |
| *sum* | 21.17±7.71 | 20.42±9.54 | 18.33±7.21 | 19.98±8.21 | 1.904@ | 0.159@ |
| *t* | 0.200 | -0.194 | 1.227 | 0.592 | 1.139# | 0.324# |
| *p* | 0.843 | 0.847 | 0.230 | 0.555 | | |

@：The *F* statistic and *P* values of main effect ；#：The *F* statistic and *P* values of interaction effect.
*：*P*<0.05，the difference was statistically significant；**：*P*<0.01，the difference was significant statistical significance.

# Figures

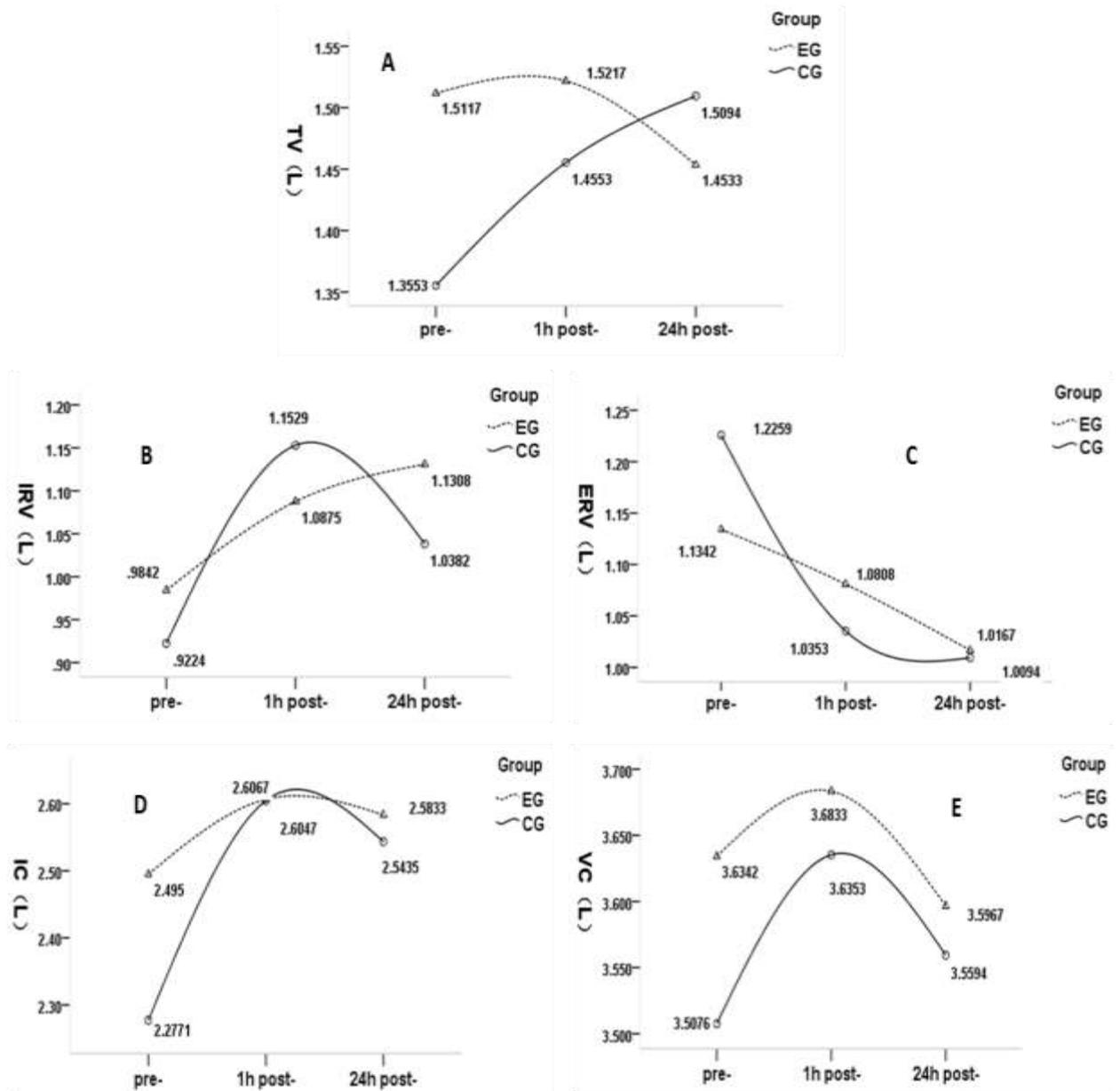

Fig. 1 Diagram of curves of static lung volume parameters of pre-hyperbaric exposure and post-hyperbaric exposure in EG & CG.
**EG:** Experimental Group; **CG:** Control Group;
**Pre-:** pre-hyperbaric exposure; **1 h post-:** 1 hour post-hyperbaric; **24 h post-:** 1 hour post-hyperbaric
【A】*TV: Tidal Volume(L)*; 【B】*IRV: Inspiratory Reserve Volume(L)*; 【C】*ERV: Expiratory Reserve Volume(L)*; 【D】*IC: Inspiratory Capacity(L)*; 【E】*VC: Vital Capacity (L)*

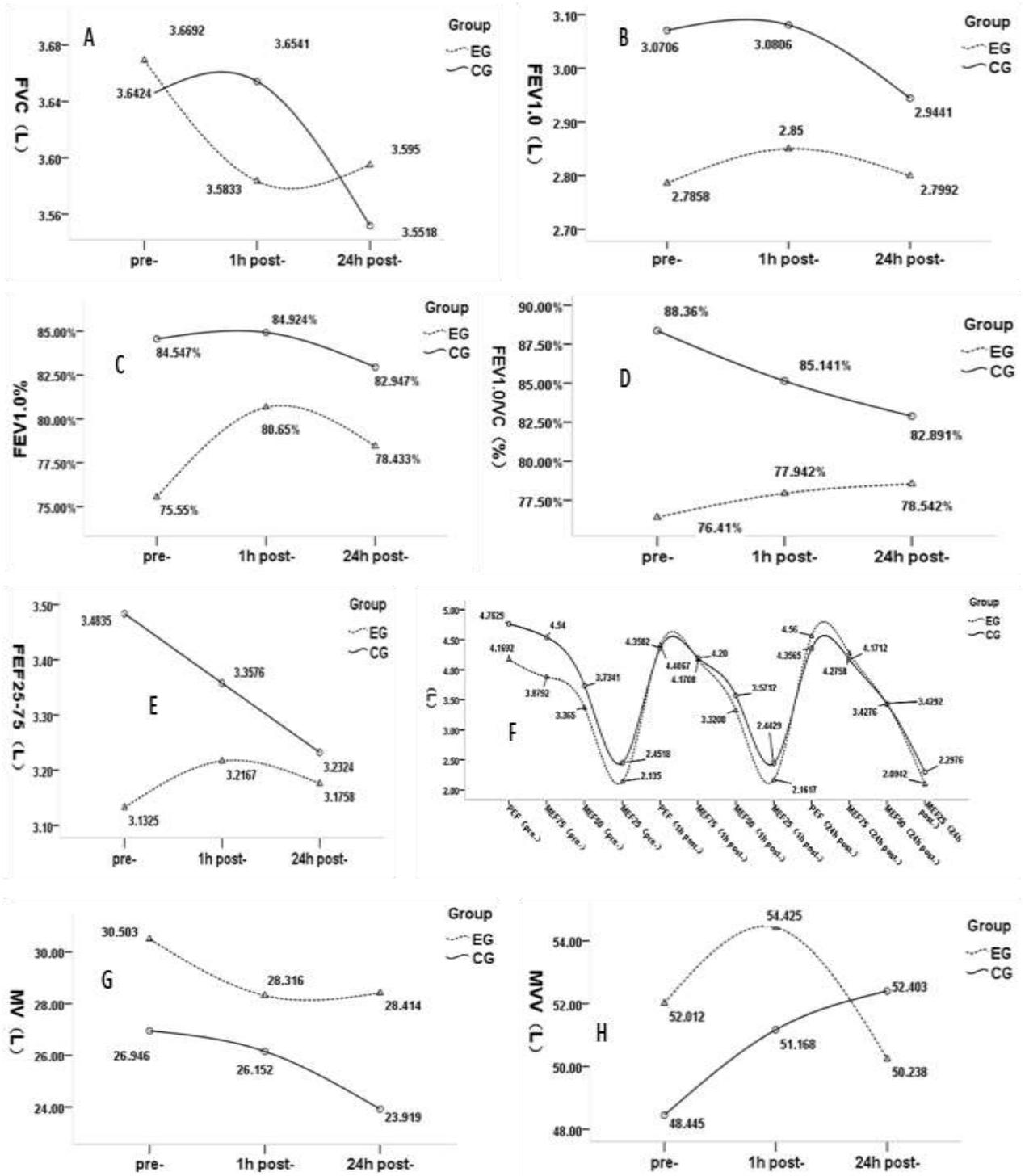

Fig. 2 Diagram of curves of dynamic lung capacity parameters of pre-hyperbaric exposure and post-hyperbaric exposure in EG & CG.

**EG:** Experimental Group; **CG:** Control Group;
**Pre-:** pre-hyperbaric exposure; **1h post-:** 1 hour post-hyperbaric; **24 h post-:** 1 hour post-hyperbaric
【A】 *FVC: Forced Vital Capacity(L)*; 【B】 *$FEV_{1.0}$: Forced Expiratory Volume in one second(L)*; 【C】 *$FEV_{1.0}$%: Ratio of $FEV_{1.0}$ to FVC*; 【D】 *$FEV_{1.0}$/VC%: Ratio of $FEV_{1.0}$ to VC*; 【E】 *$FEF_{25-75}$: Forced Expiratory Flow(L/s)*; 【F】 *PEF: Peak Expiratory Flow(L/s)*; *$MEF_{75}$: Forced Expiratory Flow after 25% of the FVC has been exhaled(L/s)*; *$MEF_{50}$: Forced Expiratory Flow after 50% of the FVC has been exhaled(L/s)*; *$MEF_{25}$: Forced Expiratory Flow after 75% of the FVC has been exhaled(L/s)*; 【G】 *MV: Minute Ventilation(L)*; 【H】 *MVV: Maximal Voluntary Ventilation (L)*

# Tables

Tab. 1  The level of ventilation function of the overall subjects

|  | n | $\bar{\chi}$ | SD |
|---|---|---|---|
| TV | 29 | 1.42 | 0.62 |
| IRV | 29 | 0.95 | 0.39 |
| ERV | 29 | 1.19 | 0.36 |
| $IV^1$ | 29 | 2.37 | 0.48 |
| $VC^2$ | 29 | 3.56 | 0.51 |
| FVC | 29 | 3.65 | 0.50 |
| $FEV_{1.0}$ | 29 | 2.95 | 0.74 |
| $FEV_{1.0}\%^3$ | 29 | 80.82% | 17.46% |
| $FEV_{1.0}\%t^4$ | 29 | 83.42% | 18.27% |
| PEF | 29 | 4.52 | 1.75 |
| $FEF_{25-75}$ | 29 | 3.34 | 1.23 |
| $MEF_{75}$ | 29 | 4.27 | 1.72 |
| $MEF_{50}$ | 29 | 3.58 | 1.34 |
| $MEF_{25}$ | 29 | 2.32 | 0.80 |
| $MV^5$ | 29 | 28.42 | 12.32 |

$^1$: IV= TV+ IRV; $^2$: VC= IRV+ TV+ ERV; $^3$: $FEV1.0\%=FEV_{1.0}/FVC\times100\%$; $^4$: $FEV1.0\%t=FEV_{1.0}/VC\times100\%$; $^5$: $MV=TV\times RR$.

Tab. 2  Changes in *TV* pre- and post- hyperbaric exposure between 2 groups

| Group | TV ($\bar{X} \pm SD$) | | | sum | F | p |
|---|---|---|---|---|---|---|
| | Pre- | 1h post- | 24h post- | | | |
| EG | 1.51±0.78 | 1.52±0.75 | 1.45±0.77 | 1.50±0.74 | 0.028 | 0.972 |
| CG | 1.36±0.49 | 1.46±0.70 | 1.51±0.57 | 1.44±0.59 | 0.295 | 0.746 |
| *sum* | 1.42±0.62 | 1.48±0.71 | 1.49±0.65 | 1.46±0.65 | 0.258@ | 0.773@ |
| *t* | 0.664 | 0.244 | -0.226 | 0.390 | 0.821# | 0.446# |
| *p* | 0.513 | 0.809 | 0.823 | 0.698 | | |

@：The *F* statistic and *P* values of main effect ；#：The *F* statistic and *P* values of interaction effect.
\*：$P<0.05$，the difference was statistically significant；\*\*：$P<0.01$，the difference was significant statistical significance.

Tab. 3  Changes in *IRV* pre- and post- hyperbaric exposure between 2 groups

| Group | IRV ($\bar{X} \pm SD$) | | | sum | F | p |
|---|---|---|---|---|---|---|
| | Pre- | 1h post- | 24h post- | | | |
| EG | 0.98±0.48 | 1.09±0.50 | 1.13±0.37 | 1.07±0.45 | 0.326 | 0.724 |
| CG | 0.92±0.32 | 1.15±0.49 | 1.04±0.40 | 1.04±0.41 | 1.343 | 0.271 |
| *sum* | 0.95±0.39 | 1.13±0.49 | 1.08±0.39 | 1.05±0.42 | 3.787@ | 0.029@* |
| *t* | 0.414 | -0.351 | 0.626 | 0.318 | 0.860# | 0.429# |
| *p* | 0.682 | 0.728 | 0.536 | 0.751 | | |

@：The *F* statistic and *P* values of main effect ；#：The *F* statistic and *P* values of interaction effect.
*：$P<0.05$，the difference was statistically significant；**：$P<0.01$，the difference was significant statistical significance.

Tab. 4  Changes in *ERV* pre- and post- hyperbaric exposure between 2 groups

| Group | *ERV* ($\bar{X} \pm SD$) | | | sum | F | p |
|---|---|---|---|---|---|---|
| | Pre- | 1h post- | 24h post- | | | |
| EG | 1.13±0.39 | 1.08±0.36 | 1.02±0.38 | 1.08±0.37 | 0.296 | 0.746 |
| CG | 1.23±0.35 | 1.04±0.37 | 1.01±0.36 | 1.09±0.37 | 1.840 | 0.170 |
| *sum* | 1.19±0.36 | 1.05±0.36 | 1.01±0.36 | 1.08±0.36 | 4.910@ | 0.011@** |
| *t* | -0.668 | 0.332 | 0.052 | -0.163 | 0.825# | 0.444# |
| *p* | 0.510 | 0.743 | 0.959 | 0.871 | | |

@：The *F* statistic and *P* values of main effect ；#：The *F* statistic and *P* values of interaction effect.
*：*P*<0.05，the difference was statistically significant；**：*P*<0.01，the difference was significant statistical significance.

Tab. 5  Changes in *IC* pre- and post- hyperbaric exposure between 2 groups

| Group | *IC* ($\bar{X} \pm SD$) | | | sum | F | p |
|---|---|---|---|---|---|---|
| | Pre- | 1h post- | 24h post- | | | |
| EG | 2.50±0.61 | 2.61±0.51 | 2.58±0.60 | 2.56±0.56 | 0.125 | 0.883 |
| CG | 2.28±0.35 | 2.60±0.50 | 2.54±0.41 | 2.47±0.44 | 2.844 | 0.068 |
| *sum* | 2.37±0.48 | 2.61±0.50 | 2.56±0.49 | 2.51±0.49 | 8.085@ | 0.001@** |
| *t* | 1.221 | 0.010 | 0.212 | 0.803 | 1.979# | 0.148# |
| *p* | 0.233 | 0.992 | 0.834 | 0.424 | | |

@：The *F* statistic and *P* values of main effect ；#：The *F* statistic and *P* values of interaction effect.
*：*P*<0.05，the difference was statistically significant；**：*P*<0.01，the difference was significant statistical significance.

Tab. 6  Changes in *VC* pre- and post- hyperbaric exposure between 2 groups

| Group | VC ($\bar{X} \pm SD$) | | | sum | F | p |
|---|---|---|---|---|---|---|
| | Pre- | 1h post- | 24h post- | | | |
| EG | 3.63±0.54 | 3.68±0.46 | 3.59±0.55 | 3.64±0.50 | 0.085 | 0.919 |
| CG | 3.51±0.51 | 3.63±0.54 | 3.56±0.54 | 3.58±0.52 | 0.251 | 0.779 |
| sum | 3.56±0.51 | 3.66±0.50 | 3.57±0.54 | 3.60±0.51 | 4.078@ | 0.022@* |
| t | 0.647 | 0.251 | 0.181 | 0.631 | 1.003# | 0.373# |
| p | 0.523 | 0.804 | 0.858 | 0.530 | | |

@：The *F* statistic and *P* values of main effect ； #：The *F* statistic and *P* values of interaction effect.
\*：$P<0.05$，the difference was statistically significant；\*\*：$P<0.01$，the difference was significant statistical significance.

Tab. 7 Changes in $FEV_{1.0}/VC$ pre- and post- hyperbaric exposure between 2 groups

| Group | $FEV1.0/VC$ ($\bar{X} \pm SD$) | | | sum | F | p |
|---|---|---|---|---|---|---|
| | Pre- | 1h post- | 24h post- | | | |
| EG | 76.41% ±22.96% | 77.94% ±15.81% | 78.54% ±19.10% | 77.63% ±18.97% | 0.038 | 0.963 |
| CG | 88.36% ±12.61% | 85.14% ±10.08% | 82.89% ±13.10% | 85.46% ±11.98% | 0.892 | 0.416 |
| *sum* | 83.42% ±18.27% | 82.16% ±13.01% | 81.09% ±15.69% | 82.22% ±15.65% | 0.197@ | 0.744@ |
| *t* | -1.637 | -1.500 | -0.729 | -2.189 | 1.046# | 0.339# |
| *p* | 0.121 | 0.145 | 0.472 | 0.033* | | |

@：The *F* statistic and *P* values of main effect； #：The *F* statistic and *P* values of interaction effect.